\documentclass[12pt]{article}
\usepackage{psfig}
\usepackage{epsf}
\def\ben{\begin{enumerate}}
\def\een{\end{enumerate}}
\def\bit{\begin{itemize}}
\def\eit{\end{itemize}}
\def\beq{\begin{equation}}
\def\eeq{\end{equation}}
\def\ba{\begin{array}}
\def\ea{\end{array}}
\def\bea{\begin{eqnarray}}
\def\eea{\end{eqnarray}}
\def\bq{\begin{quote}}
\def\eq{\end{quote}}

\def\bib{\bibitem}
\def\be{\begin{equation}}
\def\ee{\end{equation}}
\def\barr{\begin{array}}
\def\earr{\end{array}}

\def\eg{ {\em e.g.,\ }}

\def\etal{ {\em et al.}}

\def\be {\begin{equation}}
\def\ee {\end{equation}}
\def\ba {\begin{array}}
\def\ea {\end{array}}
\def\bea {\begin{eqnarray}}
\def\eea {\end{eqnarray}}

\def\r {\right}
\def\l {\left}

\def\wt {\widetilde}

\def\bc {\begin{center}}
\def\ec {\end{center}}

\def \m3{\left |{m_{3}}\right |}

\def \H2{\left |{H_{2}}\right |}
\def \MHU{m_{H_u}^2}
\def \MHD{m_{H_d}^2}
\def \MHF{m_{1/2}}
\def \MSX{m_{16}}
\def \MTN{m_{10}}
\def \MGL{m_{\tilde g}}
\def \MSQ{m_{\tilde q}}
\def \MSL{m_{\tilde l}}
\def\lapp{\mathrel{\rlap{\raise.5ex\hbox{$<$}}
                    {\lower.5ex\hbox{$\sim$}}}}
\def\gapp{\mathrel{\rlap{\raise.5ex\hbox{$>$}}
                    {\lower.5ex\hbox{$\sim$}}}}
\def\issue(#1,#2,#3){{\bf #1}, #2 (#3)} 

\def\opcit(#1){ {\em op. cit.}, #1}

\def\APP(#1,#2,#3){Acta Phys.\ Polon.\ \issue(#1,#2,#3)}
\def\ARNPS(#1,#2,#3){Ann.\ Rev.\ Nucl.\ Part.\ Sci.\ \issue(#1,#2,#3)}
\def\CPC(#1,#2,#3){Comp.\ Phys.\ Comm.\ \issue(#1,#2,#3)}
\def\CIP(#1,#2,#3){Comput.\ Phys.\ \issue(#1,#2,#3)}
\def\EPJC(#1,#2,#3){Eur.\ Phys.\ J.\ C\ \issue(#1,#2,#3)}
\def\EPJD(#1,#2,#3){Eur.\ Phys.\ J. Direct\ C\ \issue(#1,#2,#3)}
\def\IEEETNS(#1,#2,#3){IEEE Trans.\ Nucl.\ Sci.\ \issue(#1,#2,#3)}
\def\IJMP(#1,#2,#3){Int.\ J.\ Mod.\ Phys. \issue(#1,#2,#3)}
\def\MPL(#1,#2,#3){Mod.\ Phys.\ Lett.\ \issue(#1,#2,#3)}
\def\NP(#1,#2,#3){Nucl.\ Phys.\ \issue(#1,#2,#3)}
\def\NIM(#1,#2,#3){Nucl.\ Instrum.\ Meth.\ \issue(#1,#2,#3)}
\def\PL(#1,#2,#3){Phys.\ Lett.\ \issue(#1,#2,#3)}
\def\PRD(#1,#2,#3){Phys.\ Rev.\ D \issue(#1,#2,#3)}
\def\PRL(#1,#2,#3){Phys.\ Rev.\ Lett.\ \issue(#1,#2,#3)}
\def\SJNP(#1,#2,#3){Sov.\ J. Nucl.\ Phys.\ \issue(#1,#2,#3)}
\def\ZPC(#1,#2,#3){Zeit.\ Phys.\ C \issue(#1,#2,#3)}
\def\JHEP(#1,#2,#3){JHEP\ \issue(#1,#2,#3)}

\begin{document}
\title{
\vspace*{-.9truein}
Effects of the SO(10) D-Term on Yukawa Unification and Unstable Minima of 
the Supersymmetric Scalar Potential}
\author{ {\sl Amitava Datta}
\thanks{Electronic address: adatta@juphys.ernet.in}
~~and
{\sl Abhijit Samanta}
\thanks{Electronic address: abhijit@juphys.ernet.in}\\
Department of Physics, Jadavpur University, Kolkata - 700 032, India
}
\maketitle
\begin{abstract}
We study the effects of SO(10) D-terms on the allowed parameter space 
(APS) in models with  $t - b - \tau$ and $b - \tau$ Yukawa unifiction.
The former is allowed only for moderate values of the D-term, if very 
precise ($\le$ 5\%) unification is required. Next we constrain the 
parameter space by looking for different dangerous directions where 
the scalar potential may be unbounded from below (UFB1 and UFB3). 
The common trilinear coupling $A_0$ plays a significant 
role in constraing the APS. For very  precise  $t - b - \tau$ Yukawa 
unification, $-\MSX \lapp A_0 \lapp \MSX$ can be probed at the LHC,
where $\MSX$ is the common soft breaking mass for the sfermions. 
Moreover, an interesting mass hierarchy with very heavy sfermions 
but light gauginos, which is strongly disfavoured in models without 
D-terms, becomes fairly common in the presence of the D-terms. The APS 
exhibits interesting characteristics if $\MSX$ is not the same as the 
soft breaking mass $\MTN$ for the Higgs sector.
In $b - \tau$ unification models with D-terms, the APS consistent with
Yukawa unification and radiative electroweak symmetry breaking, 
increases as the UFB1 constraint becomes weaker. However for $A_0 \lapp 0$, 
a stronger UFB3 condition still puts, for a given $\MSX$, a stringent 
upper bound on the common gaugino mass ($\MHF$)  and a  lower bound on  
$\MSX$ for a given $\MHF$. The effects of sign
of $\mu$ on Yukawa unification and UFB constraints are also discussed. 
\end{abstract}
\setcounter{footnote}{0}
\renewcommand{\thefootnote}{\arabic{footnote}}


\section{Introduction}

It is quite possible that the Standard Model (SM), is not the 
ultimate theory of nature, as is hinted by a number of theoretical 
shortcomings. One of the
most popular choices for physics beyond SM is supersymmetry (SUSY)
\cite{susy-review}. However, the experimental requirement that SUSY 
must be a broken symmetry introduces a plethora of new soft breaking parameters.
There are important constraints on this large parameter space from 
the negative results of the sparticle searches at colliders like
LEP\cite{lepbound} and Tevatron\cite{tevatronbound}. In addition there 
are important theoretical constraints which are often introduced 
for aesthetic reasons. From practical point of view, however, the 
most important effect of such constraints is to reduce the number of
free parameters. For example, the assumption that the soft breaking
terms arise as a result of gravitational interactions leads to the popular 
minimal supergravity (mSUGRA) model with five free parameters only, 
defined at a high energy scale where SUSY is broken.
They are the common scalar mass ($m_0$), the common gaugino mass ($\MHF$),
the common trilinear coupling ($A_0$), the ratio of vacuum expectation 
values of two Higgs field ($\tan\beta$) and the sign of $\mu$, the
higgsino mass parameter.
In this paper we shall restrict ourselves to variations of this
basic framework.

A very useful way to further constrain the allowed parameter space (APS) 
of softly broken SUSY
models is to consider the dangerous directions of the scalar potential,
where the potential  may be unbounded from below (UFB) or develop
a charge and/or color breaking (CCB) minima \cite{oldufb}. 
Different directions are chosen by
giving vacuum expectation value (VEV) to one or more  
coloured and / or charged scalar fields, while
the VEVs of the other scalars are taken to be zero.

In a very interesting  paper which revived interest in
UFB and CCB constraints, Casas {\em et al} \cite{casas} investigated
the effects of such constraints on SUSY models. Though their
formulae  are fairly model-independent, they had carried out
the numerical analysis within the framework of mSUGRA for
 moderate values  of $\tan\beta$ only, when one can ignore the
effects of b and $\tau$ Yukawa couplings in the relevant
renormalization group equations (RGEs).
Their main result was that  a
certain
UFB constraint known as UFB3 with VEVs given in the direction of the
slepton fields  puts the tightest bound on the SUSY parameter
space that they considered (see eq. (93) of \cite{casas} and the discussions
that follow).

In an earlier paper \cite{paper1}, we had extended and complemented the
work of \cite{casas} by looking at the APS
subject to such `potential constraints'
for large values of $\tan\beta$, motivated by partial $b$-$\tau$ 
\cite{b-tau,partial} or full
$t$-$b$-$\tau$ Yukawa unification \cite{sotenbr}. Such unifications are
natural consequences of an underlying Grand Unified Theory 
(GUT). We considered a popular model  in which the GUT
group SO(10) breaks directly into the SM gauge group SU(3) $\times$ SU(2)
$\times$ U(1). All matter fields belonging to a particular generation 
is contained in a 16 dimensional representation of SO(10). 
With a minimal Higgs field content (one {\bf 10}-plet
containing both the Higgs doublets required to give masses to u and d
type quarks) 
all three Yukawa couplings related to the third generation fermions 
must unify at the GUT scale. If one assumes
more than one {\bf 10}-plet, at least the bottom and the tau Yukawa
couplings should unify.

In \cite{paper1} we assumed 
a common soft breaking (SB) mass ($\MSX$) at $M_G$   for all sfermions
of a given generation. Similarly a common
mass parameter ($\MTN$) was chosen  for both the Higgs fields. We then 
studied the stability of the potential for 
two sets of boundary conditions: i)
 the mSUGRA motivated universal scenario ($\MSX = \MTN$),
and ii) a  nonuniversal scenario ($\MSX \neq \MTN$). The second condition 
 is motivated by the fact that a common scalar mass at the
Planck scale, generated, \eg by the SUGRA mechanism,  may lead to nonuniversal
scalar masses at $M_G$ due
to different running of $m_{10}$ and $m_{16}$, as they belong to
different GUT multiplets \cite{running}.

In this paper we shall extend the work of \cite{paper1} by considering 
the APS due to Yukawa unification and UFB constraints in the presence
of SO(10) breaking D-terms. The group $SO(10)$ contains $SU(5)\times
U(1)_X$ as a subgroup.  
It is well known that the breaking of  SO(10) to the  lower rank SM group 
may introduce nonzero D-terms at the GUT scale\cite{d-term}.
We further assume that the D-terms are linked to
the breaking of $U(1)_X$ only. 
It should be noted that if one assumes the existence of additional $U(1)$'s
at high energies, it is quite natural to assume that the D-term contributions
to scalar masses are non-zero\cite{d-term}. The only uncertainty
lies in the magnitude of the D-terms which may or may not be significant.
The squark - slepton and Higgs soft breaking masses in this case can be parametrized as 
\bc
$m_{\tilde Q}^2 = m_{\tilde E}^2 = m_{\tilde U}^2 = m_{16}^2 +
m_D^2$\\ $m_{\tilde D}^2 =m_{\tilde L}^2 =m_{16}^2 -3 m_D^2$\\
$m_{H_{d,u}}^2 =m_{10}^2 \pm 2 m_D^2$\\
\ec
where $\tilde Q$ and $\tilde L$ are SU(2) doublets of
squarks and sleptons, 
$\tilde E$, $\tilde U$ and $\tilde D$ are  SU(2) singlet
 charged sleptons, up and down type squarks respectively. 
The unknown parameter $m_D^2$ (the D-term)  can be of either
sign. The mass differences
arise because of the differences in the $U(1)$ quantum numbers of the
sparticles concerned.  As can be readily seen from the above formula for
$m_D^2>$ 0, the left handed sleptons  and right handed down type
squarks (belonging to the $\bar 5$ representation of $SU(5)$),
are lighter than the members of the $10$ plet of $SU(5)$.
In recent times the phenomenology of
the D-terms has attained wide attention\cite{tata,pheno-d}.

D-terms acquire particular significance in the context of $t-b-\tau$ 
unification as has already been noted in the literature \cite{tata}.
A new result of this work is that while moderate values of D-terms 
facilitate very accurate unification, high values of this parameter spoil it.

The UFB and CCB constraints depend crucially on the particle spectra
at the properly chosen scale  where the true minimum and the dangerous 
minimum can be reliably evaluated from the tree level potential 
($V_{tree}$) \cite{casas,gamberini}. 
Such spectra, in turn, depend on the boundary conditions at
the GUT scale.
The SO(10) breaking D-terms alter the sparticle spectra at the GUT
scale  and may affect the stability of the potential. 
In this paper we focus our attention on  the impact of such 
D-terms on the APS restricted by  Yukawa
unification and  the stability of the potential in both universal
and nonuniversal scenarios. 


Throughout the paper we ignore the possibility that nonrenormalizable
effective operators may stabilise the potential \cite{nonrenorm}. The
dangerous minima that we encounter in our analysis typically occur at
scales $\lapp 10^8$ GeV where the effects induced by the 
nonrenormalizable operators, which in principle can be significant
in the vicinity of the GUT scale,  are not likely to be very serious.

It has  been pointed out in the literature
that the standard vacuum, though
metastable, may have a lifetime longer than the age of the universe
\cite{claudson}, while the true vacuum is indeed charge and colour
breaking.
If this  be the case, the theory seems to be
acceptable in spite of the existence of the unacceptable UFB minima that
we have analysed. However, the life-time calculation, which is relatively
straightforward for a single scalar field, is much more uncertain in
theories where the potential is a function of many scalar fields. Thus
it is difficult to judge the reliability of these calculations. Moreover,
the constraints obtained by us does not loose their significance even
if the false vacuum idea  happens to be the correct theory. If these
constraints are violated by future expeimental data then that would
automatically lead to the startling conclusion that  we are living in
a false vacuum  and charge and colour symmetry may eventually breakdown.

It has been known for quite some time  that while $\mu > 0$(in our
sign convention which is opposite to that of Haber and Kane
\cite{susy-review}) is required by Yukawa unification,
the opposite sign is preferred by the data on the branching ratio
of b $\rightarrow$ s $\gamma$ and that on $g_\mu$ - 2 (
see \cite{ferrandis,referee} for some of the recent analyses and
references to the earlier works).

It has been shown in \cite{ferrandis} and also in the first paper of
\cite{referee} that in a narrow region of the parameter space
there
is no conflict between data and Yukawa unification. We have analysed the
parameter space found in \cite{ferrandis} in the light of the stability
of the vacuum  and the results are given in the next section
(see Table 1 in particular). The
above conflict may also be resolved by introducing non-universal gaugino
masses ( see Chattopadhaya and Nath in \cite{referee}).

In section 2 we discuss the effects of $\tan\beta$, $m_D$
and sign of $\mu$ on Yukawa unification and stability of the potential.
In subsection 2.2 and 2.3 we study the APS for both $t-b-\tau$ 
and $b-\tau$ unification in conjunction with the UFB constraints.
In the last section we summarise and conclude.  

\section{Results}
\subsection{General Discussions}
The methodology of finding the spectra is the same as in \cite{paper1},
which is based on the computer program ISASUGRA, a part of the ISAJET
package, vesion 7.48\cite{ISAJET}.  The parameters
$\mu$ and $B$ are fixed by radiative electroweak symmetry breaking 
(REWSB) \cite{rewsb} at a scale $M_S = \sqrt{m_{\tilde t_L} m_{\tilde t_R}}$. 
We further require that the lightest neutralino
($\tilde \chi^0$) be the lightest supersymmetric particle (LSP). 
The above two constraints will also be used to obtain the allowed parameter space (APS)
although their use may not  be mentioned explicitly everywhere.
We then fix $\tan\beta$ to its lowest value required by Yukawa unification. 
Next  we check the  experimental constraints on sparticle masses. 
Finally we impose the UFB constraints.

Before discussing the basic reasons of how Yukawa unification 
plays a significant role in restricting the APS, we  will review the different uncertainties
of Yukawa unification.
The effectiveness of Yukawa unification as a restrictor of the APS
diminishes, as expected, as the accuracy with
which we require the unification to hold good is relaxed.
There are several reasons why the unification may not be exact. 
First, there may be threshold corrections \cite{threshold}, both at the 
SUSY breaking scale (due to nondegeneracy of the sparticles) and at $M_G$,
of which no exact estimate exist. Secondly,
we have used two-loop RGEs for the evolutions of gauge
couplings as well as Yukawa couplings and one loop RGEs for the soft 
breaking parameters, but higher order loop corrections may be important 
at a few percent level at higher energy scales. 
Finally the success of the unification program is also dependent 
on the choice of $\alpha_s(M_Z)$  
which is not known as precisely as $\alpha_1$ or $\alpha_2$.
To take into account such uncertainties, one relaxes the Yukawa 
unification condition
to a finite amount (5\%, 10\% or 20\%) which should indirectly
take care of the above caveats.  
The demand of very accurate Yukawa coupling unification at $M_G$
puts severe constraint on $\tan\beta$ restricting it to very large 
values only. 

The accuracy of the  $t-b-\tau$ unification
is usually relaxed since there are more elements of uncertainty, {\em e.g.}, the
choice of the Higgs sector. To quantify this accuracy,
one can  define three variables $r_{b\tau}, r_{tb}$ 
and $r_{t\tau}$ where generically $r_{xy}=Max(Y_x/Y_y,Y_y/Y_x)$. 
For example, to check whether the couplings unify, one should select 
only those points in the parameter space where, \eg
$Max(r_{b\tau},r_{tb},r_{t\tau})<1.10$
(for 10\% $t-b-\tau$ unification) and $r_{b\tau}<1.05$ 
(for 5\% $b-\tau$ unification). 

Now we will focus on the basic reasons which lead to upper and lower
bounds on the APS in the $\MSX-\MHF$ plane, 
if partial ($b-\tau$) or full ($t-b-\tau$) unification is required.
It is wellknown that for precise Yukawa unification 
one should have $\mu >0$.
The partial Yukawa unification can be accommodated at relatively low
values of $\tan\beta$ when the phenomenologically interesting  small 
$\MSX,\MHF$ region of the parameter space is allowed
(viz. for $\MSX,\MHF\sim$ 200 GeV, the required minimum value of
tan$\beta\sim 30,$ and for $\MSX,\MHF\sim$ 800 GeV, 
${(\tan\beta)}_{\rm min}\sim$ 41).
On the otherhand  $\tan\beta$ cannot be arbitrarily increased
due to the REWSB. This basic trend, which often makes the two constraints 
incompatible, remains unaltered irrespective of the choice of the other 
parameters. 

The constraints
due to Yukawa unification and REWSB are relatively weak for large
negative values of $A_0$ and becomes stronger
as this parameter is algebraically increased 
\footnote{This is due to the fact that unification holds at 
relatively lower values of $\tan\beta$ as one goes to larger 
negative values of $A_0$. There is, therefore, more room for
increasing $\tan\beta$, if required, without violating REWSB 
condition. This point was not ellaborated in our earlier 
work\cite{paper1}.}.
On the other hand, the UFB constraints are very potent
for large negative values of $A_0$. The expanded APS allowed by Yukawa 
unification, is eaten up by the UFB constraints.  In this sense the 
two sets of constraints are complementary\cite{paper1}.

$Y_t$ varies relatively slowly with respect to
tan$\beta$ compared to $Y_{\tau}$ and $Y_b$.  For very accurate (5 \%) 
$t-b-\tau$ unification, we, therefore, need high values of 
$\tan\beta\sim 47 - 51$. 
In this case the  low $\MSX - \MHF$ region is excluded by the 
REWSB condition,
leading to lower bounds much stronger than the experimental ones and 
the resulting APS is restricted to phenomenologically uninteresting
high $\MSX,\MHF$ region. For example,
with $\tan\beta$ =49.5 the lowest allowed values are 
$\MSX=$600GeV, $\MHF=1000$GeV leading to rather heavy sparticles.
                                                           
\begin{figure}[htb]
\centerline{
\psfig{file=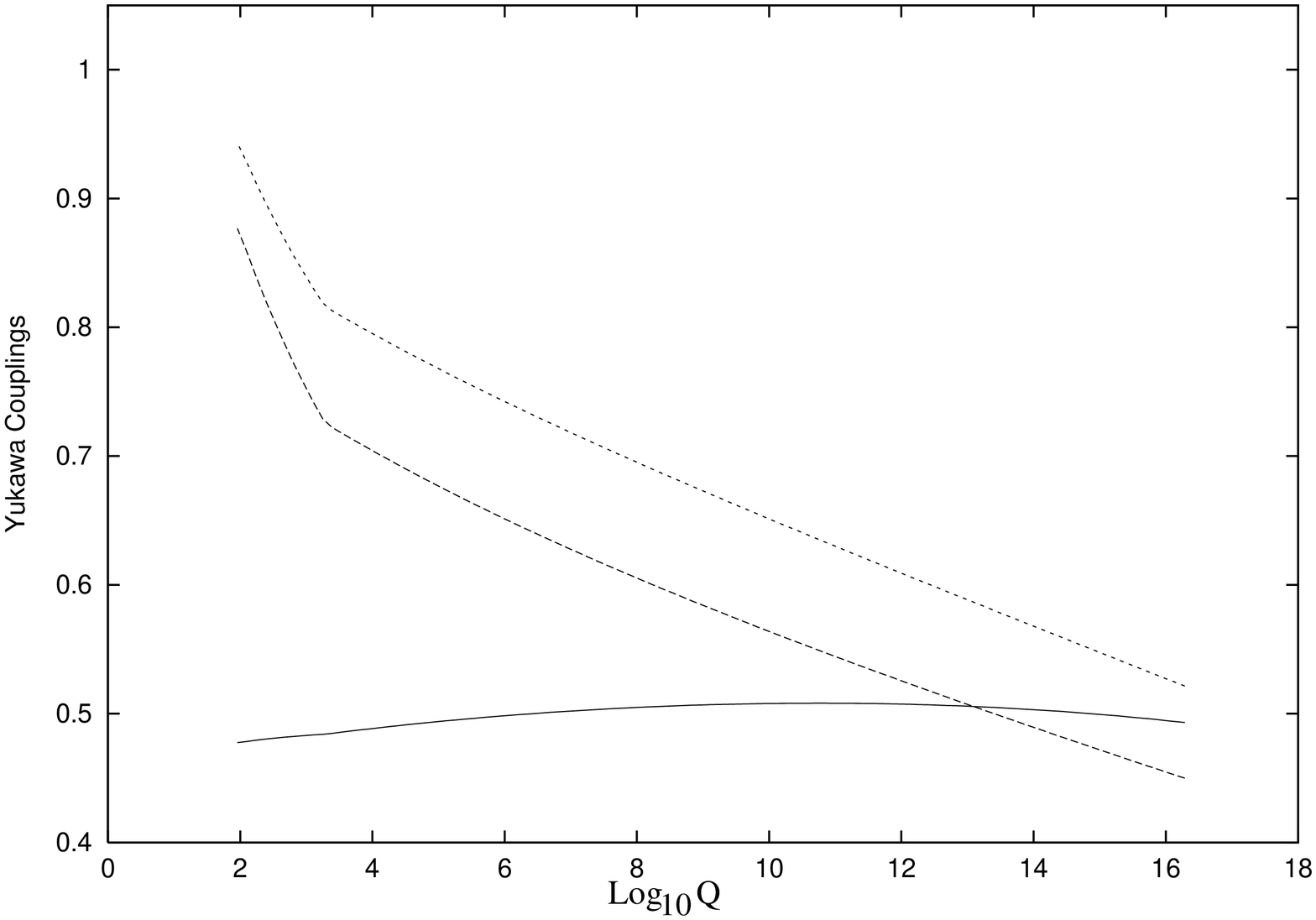,width=8cm,angle=270}}
\caption{\sl{The variation of Yukawa couplings
            with renormalization scale Q (GeV).
            From above the lines are for top, bottom and $\tau$ Yukawa
            couplings respectively.
            We have used $m_{16}=m_{10}$=1.5 TeV, $m_{1/2}=0.5$ TeV.
            $A_0=0$, $\tan\beta= 48.5$ and $m_D$ =0.
            }}
    \label{yu0}
\end{figure}
\begin{figure}
\centerline{
\psfig{file=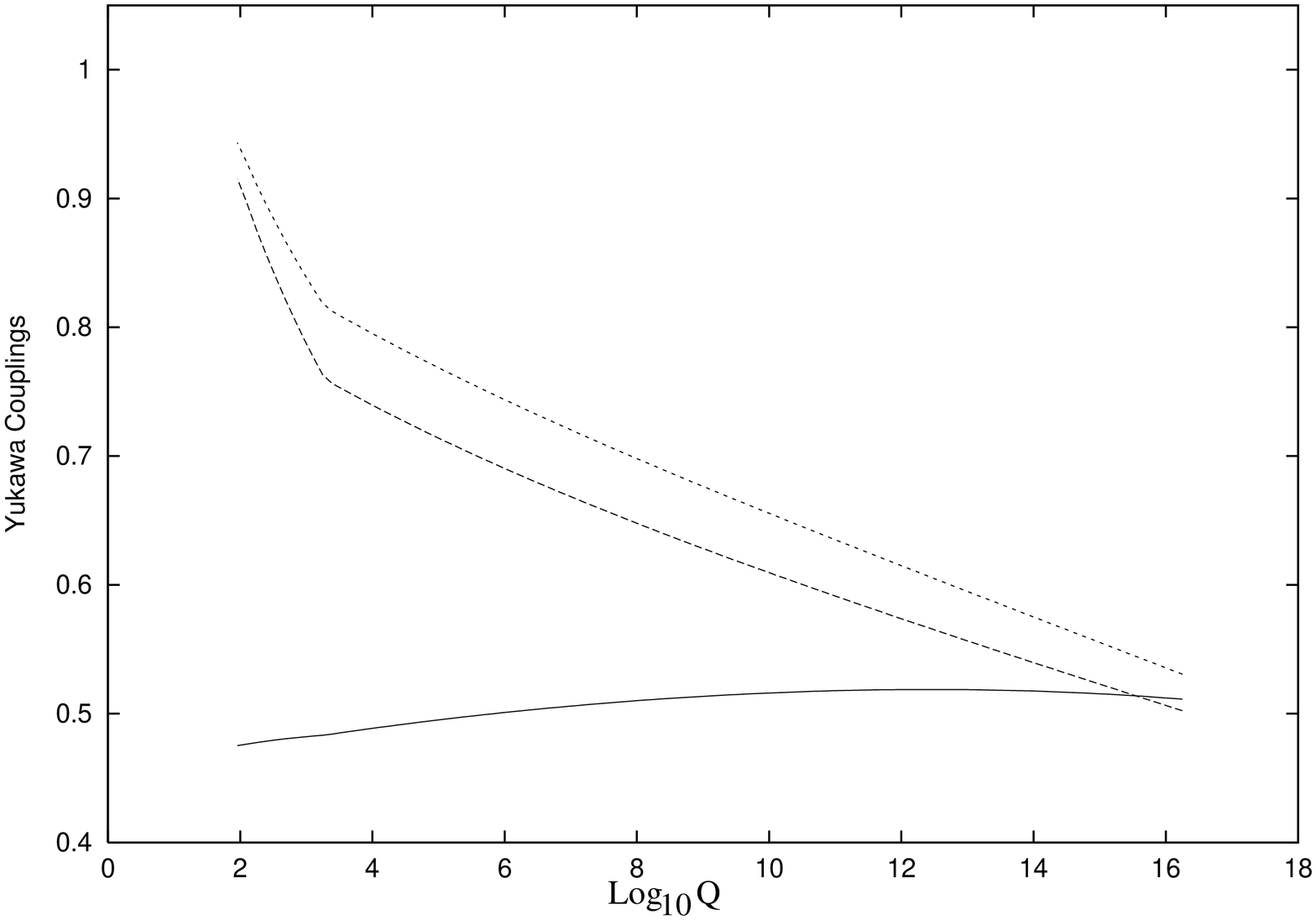,width=8cm,angle=270}}
\caption{\sl{The same as fig. \ref{yu0}, with $m_D=\MSX/5$. }}
     \label{yu5}
\end{figure}
\begin{figure}[htb]
\centerline{
\psfig{file=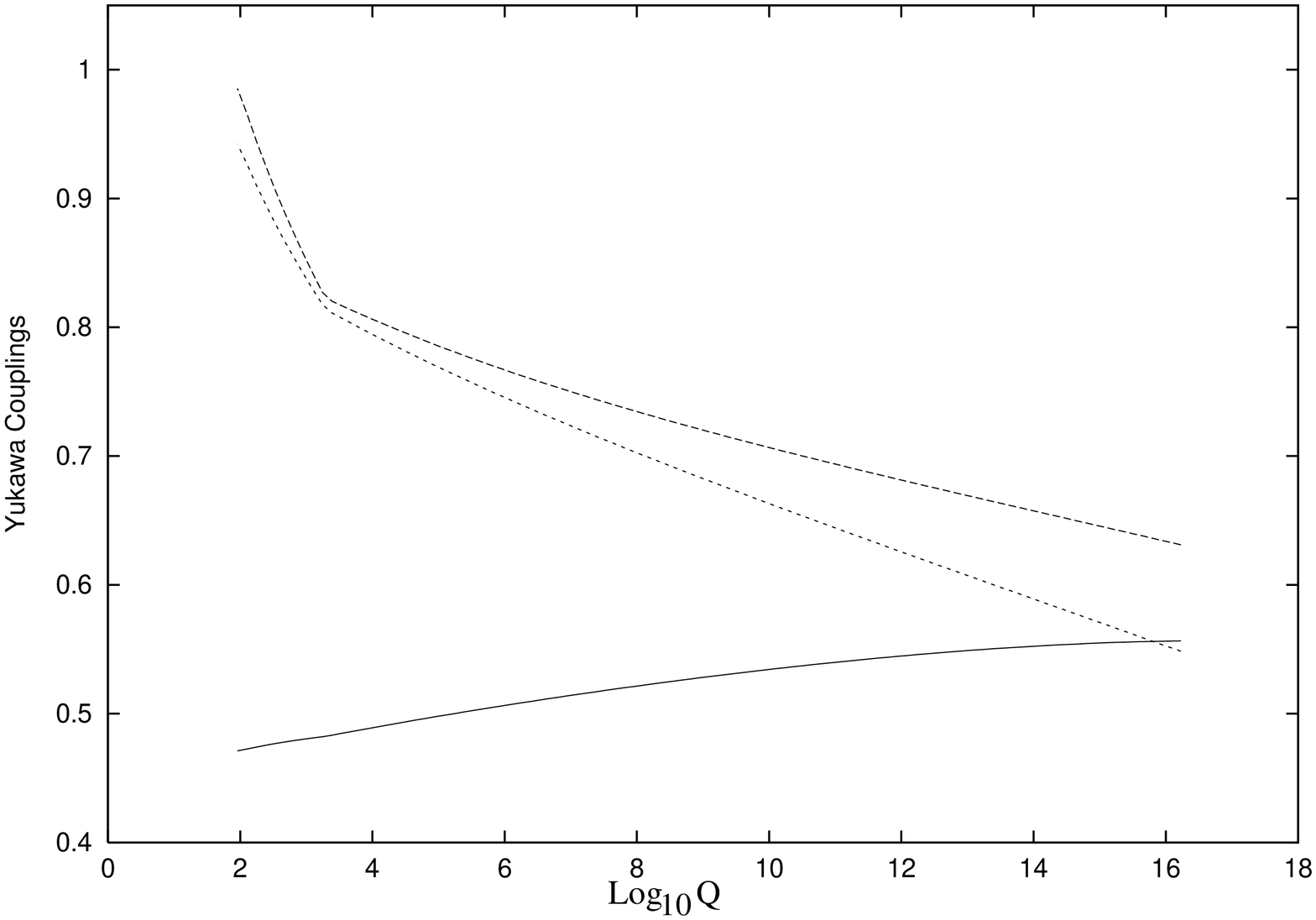,width=8cm,angle=270}
}
\caption{\sl{The same as fig. \ref{yu0}, with $m_D=\MSX/3$.
           }}
     \label{yu3}
\end{figure}

In the presence of  D-terms a larger APS is obtained even if very
accurate full unification is required\cite{tata}. 
A new finding of this paper is that though moderate values of 
$m_D$ leads to better Yukawa
unification, somewhat larger values spoil it. Although
the D-terms do not affect the evolution of the Yukawa couplings directly
through the RGEs, they change the initial
conditions through SUSY radiative corrections to $m_b(m_Z)$ \cite{pierce}.
This is illustrated in figs. \ref{yu0}---\ref{yu3}, where 
approximate unification is studied for three different values of $m_D$. 
The choice of other SUSY parameters for these figures are as follows:
\bc
$m_{10}=m_{16}$ = 1500GeV, \ $m_{1/2}$ = 500GeV, \ tan$\beta$=48.5, \
$A_0$ = 0 and $\mu >0$. 
\ec
From fig \ref{yu0} ($m_D=0$), we see that the accuracy of unification 
is rather modest ($\sim 15\%$). As $m_D$ is further increased to 
$m_{16}/5$ (fig.\
\ref{yu5}), the $\tilde b\tilde g$ loop corrections (see eq.\ (8) of
\cite{pierce}) to $m_b(m_Z)$ increases and leads to better unification.
However, if we increase $m_D$ further to $m_{16}/3$, the accuracy of 
unification deteriorates ( fig.\ \ref{yu3}) since $m_b(m_Z)$ suffers 
a correction which is too
large. We have checked that this feature holds for a wide choice of SUSY
parameters. 
%
%

\begin{figure}[htb]
\vspace*{1cm}
\centerline{
\psfig{file=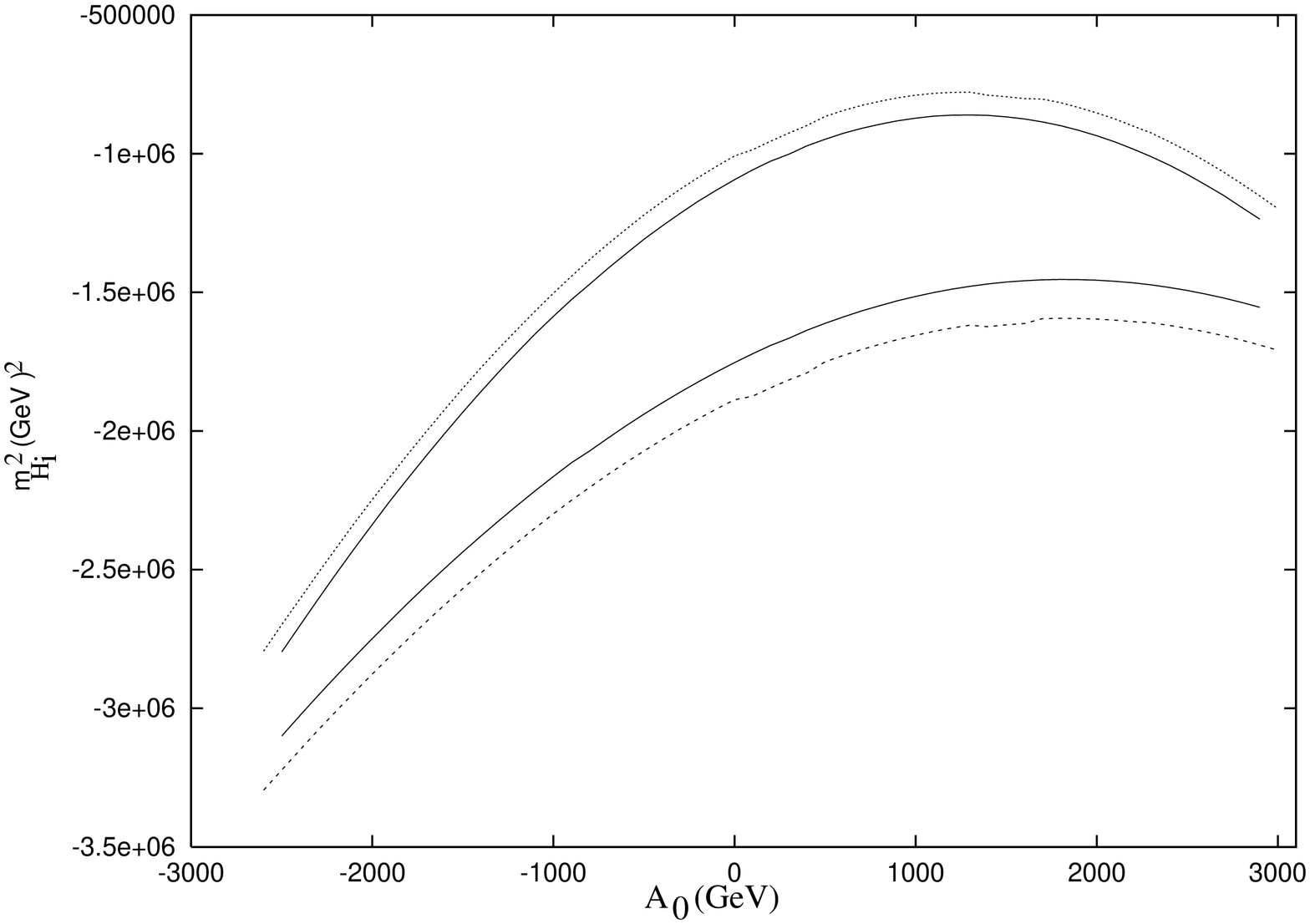,width=10cm,angle=270}
}
\caption{\sl{The variation of the Higgs mass parameters $\MHD$ and
            $\MHU$, evaluted at the scale $M_S = \sqrt{m_{t_L} m_{t_R}}$,
            with the trilinear coupling $A_0$. 
            The solid (dotted) lines are for $m_D=\MSX/5 (\MSX/3)$.
            The top two lines are for $\MHD$ while the lower pair
           is for$\MHU$. We have used $m_{16}=m_{10}=m_{1/2}=1$ TeV,
            tan$\beta$ = 45. 
            }}
    \label{mh2md}
\end{figure}

Quite often the APS expanded due to the presence of D-terms
is significantly reduced by the UFB constraints.
As discussed in our earlier work\cite{paper1},  the 
variation of $\MHU$ and $\MHD$, the soft breaking masses of the two 
Higgs bosons, with respect to the common trilinear
coupling $A_0$ is of crucial importance in understanding this.
 Here we extend the
discussion for non-zero values of the D-term, $m_D =$ $\MSX/5$ and
$\MSX/3$. The effects are  illustrated in fig. \ref{mh2md}.
As we increase the magnitude of the D-term, the UFB3 becomes more 
potent though UFB1 looses its restrictive power
for a fixed value of $\tan\beta$.
To clarify this result we examine two important expressions
of Casas \etal\cite{casas}.
The first one is  
\be
\MHU+\MHD +2\mu^2 \geq 2 \left |\mu B\right |,
    \label{ufbone}
\ee
which is known as the UFB1 condition and should be satisfied at any 
scale $\hat{Q}>M_S$, in particular at the unification scale $\hat{Q}=M_G$.
The second one is the UFB3 constraint,
\be
V_{UFB3}=[\MHU+m_{L_{i}}^{2}]\l|H_u\r|^2+\frac{\left|\mu\right|}
{\lambda_{E_j}}[m_{L_j}^{2}
+m_{E_{j}}^{2}+m_{L_{i}}^{2}]\l|H_u\r|-
\frac{2{m_{L_i}}^4}{{g^\prime}^2+g^2},
\label{ufbthree}
\ee
where $g^\prime$ and $g$ are normalised gauge couplings of $U(1)$ and
$SU(2)$ respectively, $\lambda_{E_j}$ is a Yukawa coupling and $i,j$
are generation indices.

We find that larger $m_D$ drives $\MHU$ to more negative values, while
$\MHD$ is driven to positive values (see fig. \ref{mh2md}). In addition, it follows
from REWSB condition  that as
the difference $\MHD - \MHU$ increases, the higgsino mass parameter $\mu$
increases. As a result  the UFB1 constraint becomes weaker
for large $m_D$ values (see eq. \ref{ufbone}).
On the otherhand at the GUT scale, $m_{L_i}^2$ becomes smaller for 
larger $m_D$. 
From eq. \ref{ufbthree} it can be concluded that the parameter space where
$\MHU + m_{L_i}^2$ is negative increases and the model is 
more succeptible to  the UFB3 codition. These effects will be reflected 
in $b-\tau$ unification as well. 

For precise ($\leq$ 5\%) $t-b-\tau$ unification the required
$\tan\beta$ is very high ($\sim 49$) and the allowed $\MSX$ 
values are large.
Here the magnitude of $\mu$ as determined by the REWSB becomes very low even for moderate
values of the D-term ($m_D \approx \MSX/5$). Consequently UFB1 
still disallows
a significant part of the enlarged APS obtained with 
introduction of the D-term. However, the effectiveness of the UFB1
constraint depends crucially on $A_0$.
We see that  $A_0 \lapp \MSX$ is ruled out by UFB1 if the D-term is zero.
In presence of the D-terms the UFB1 constraints 
become weaker but still have some restrictive power for $A_0 \lapp 0$.
Moreover UFB3 becomes weaker for large $\MSX$ in general.

Now we shall discuss the impact of the sign of $\mu$ on both UFB 
constraints and Yukawa unification. Yukawa unification 
generally favours $\mu > 0$. 
The sign of $\mu$ affects unification through loop corrections\cite{pierce} 
to the bottom Yukawa coupling, which are incorporated at the weak scale. 
These corrections lower the bottom Yukawa coupling significantly, 
consequently the GUT scale value bcomes very low, which tends to spoil 
Yukawa unification. Baer\etal\cite{ferrandis} showed that for $\mu < 0$
full unification with low accuracy ($\sim 30\%$) is possible.
This is interesting since approximate
unification then becomes consistent with the constraints from
$b \rightarrow s \gamma$ and $g - 2$ of the muon. 
It was shown in \cite{ferrandis} that the Yukawa unified APS 
favours $A_0 \approx -2\MSX$ and $\MTN\approx\sqrt{2}\MSX$ 
(see fig. 1 of \cite{ferrandis}).
We have extended the analysis of ref\cite{paper1} for $\mu < 0$ 
and have found  that the UFB1 condition looses it effectiveness for 
$\mu <0$. The bottom Yukawa coupling 
affects the value of $\MHD$ through renormalization group (RG) running and 
cannot make $\MHD$ large negative as in the $\mu >0$ case.
This is why UFB1 is weakened (see eq. \ref{ufbone}).
We have studied the APS obtained in \cite{ferrandis} and found that 
UFB1 can disallow certain negative values of $m_D^2$ depending on the
magnitudes of $\MSX$ and $\MHF$. 
If $\MSX, \MHF$ are increased, relatively small  negative
values  of $m_D^2$ make the potential unstable under UFB1 condition.  
Some representative regions of APS are shown in Table 1. In obtaining 
Table 1, $A_0$, $\MTN$ and $\tan\beta$ are varied within the ranges 
indicated by ref\cite{ferrandis}.
\begin{table}[htbp]
\caption{\sl{Representative D-terms allowed by UFB1 for $\mu <0$.
}}
\begin{center}
\begin{tabular}{|c|c|c|c|c|c|}
\hline
$\MSX$ & $\MHF$ & allowed $m_D^2$ \\
(GeV) & (GeV) & (GeV$^2$)\\
\hline
600 & 300 & $\gapp -(\MSX/4.3)^2$\\
1000&  300 & $\gapp -(\MSX/4.5)^2$\\
1000 & 500 & $\gapp -(\MSX/5.0)^2$\\
\hline
\end{tabular}
\end{center}
\end{table}

We have also checked that precise $b-\tau$ Yukawa unification is 
not possible for $\mu <0$ except for very low $\tan\beta(\sim 1)$. 
As $\tan\beta$ is increased, $Y_\tau$ at $M_G$ increases rapidly 
compared to $Y_b$. This is clear from 
$\wt t\wt\chi^+$ loop correction (see eqn. 15 of \cite{pierce}).

\subsection{$t$-$b$-$\tau$ Unification}
Through out this section we shall restrict ourselves to unification 
within 5\%. 
For the sake of completeness and systematic analysis, we start our discussion
for large negative values of $A_0$ (say, $A_0 = -2\MSX$), though it is  
not interesting from the point of view of collider searches.
We first consider moderate values of the D - term (\eg 
$m_D = \MSX/5$). 
Though at low values of $\tan\beta$ the large negative values of $A_0$ 
are favoured by REWSB (see, \eg the following section on 
$b-\tau$ unifiaction), they are strongly 
disfavoured at large tan$\beta$ ($\sim$ 49) which is
required by full unification. Only 
a narrow band of $\MHF$ is allowed.  However,
the APS corresponds to rather heavy sparticles (e.g., $\MSX(\MHF)
\gapp$ 1100(1300)GeV) which are of little interest  even for SUSY 
searches at the LHC. Non-universality  affects the APS marginally;
no significant change can be obtained. 
Thus no squarks - gluino 
signal is expected at LHC for $A_0 \lapp -2\MSX$ irrespective of the
boundary conditions (universal or non-universal) on  the scalar masses. 
Over a small region 
of the APS somewhat lighter  sleptons ($\MSL \sim$ 1000GeV) are still
permitted.
Moreover, the tiny APS allowed by the 
unification criterion is ruled out by the UFB1 condition.
For $A_0 \gapp 2\MSX$, the APS is qualitatively the same as that for
$A_0 \lapp -2\MSX$, with the only difference that the UFB constraint
does not play any role.

Relatively large APSs with phenomenologically interesting sparticle masses
open up
for $-\MSX \lapp A_0 \lapp \MSX$, which is favourable for both 
Yukawa unification and REWSB. The common feature of the
APS is that  gluino masses almost as low as the current 
experimental lower bound with much heavier squark and slepton 
masses($\gapp$ 1TeV) can be obtained irrespective of universality
or non-universality of scalar masses.
It should be stressed that this mass pattern 
cannot be accommodated without the D-terms. In the presence of D-term 
this mass hierarchy becomes a distinct possibility.

\begin{figure}[htb]
\centerline{
\psfig{file=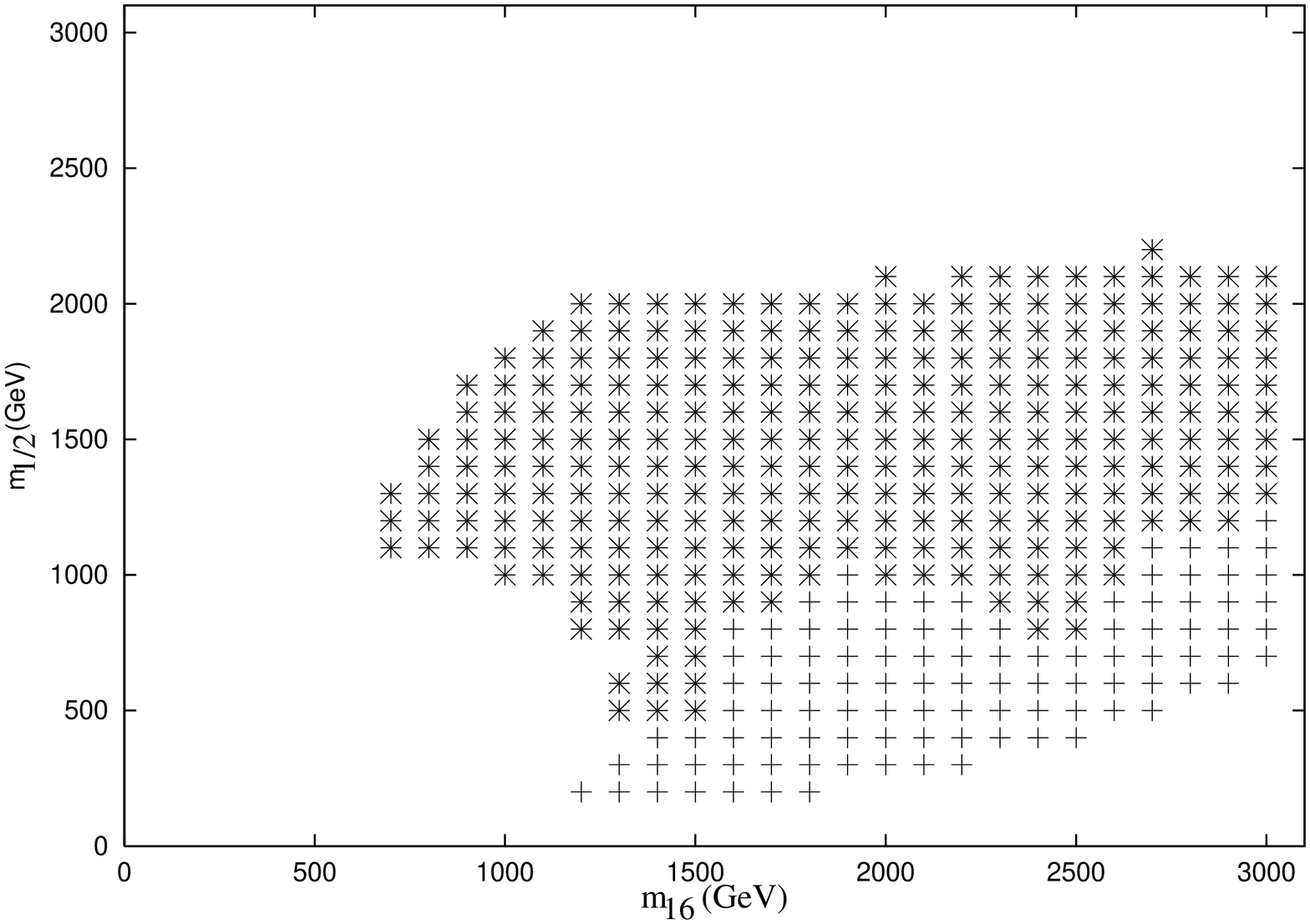,width=12cm,angle=270}
}
\caption{\sl{The allowed parameter space in the universal scenario with
             $t-b-\tau$ unification $\le$ 5\%. All the points are 
             allowed by the Yukawa unification criterion; the asterisks 
             are ruled out by UFB1.
             We set $m_D = \MSX/5$ and $A_0=-m_{16}$.
            }}
    \label{tbtaum0p1md5a0_1}
\vspace*{5mm}
\centerline{
\psfig{file=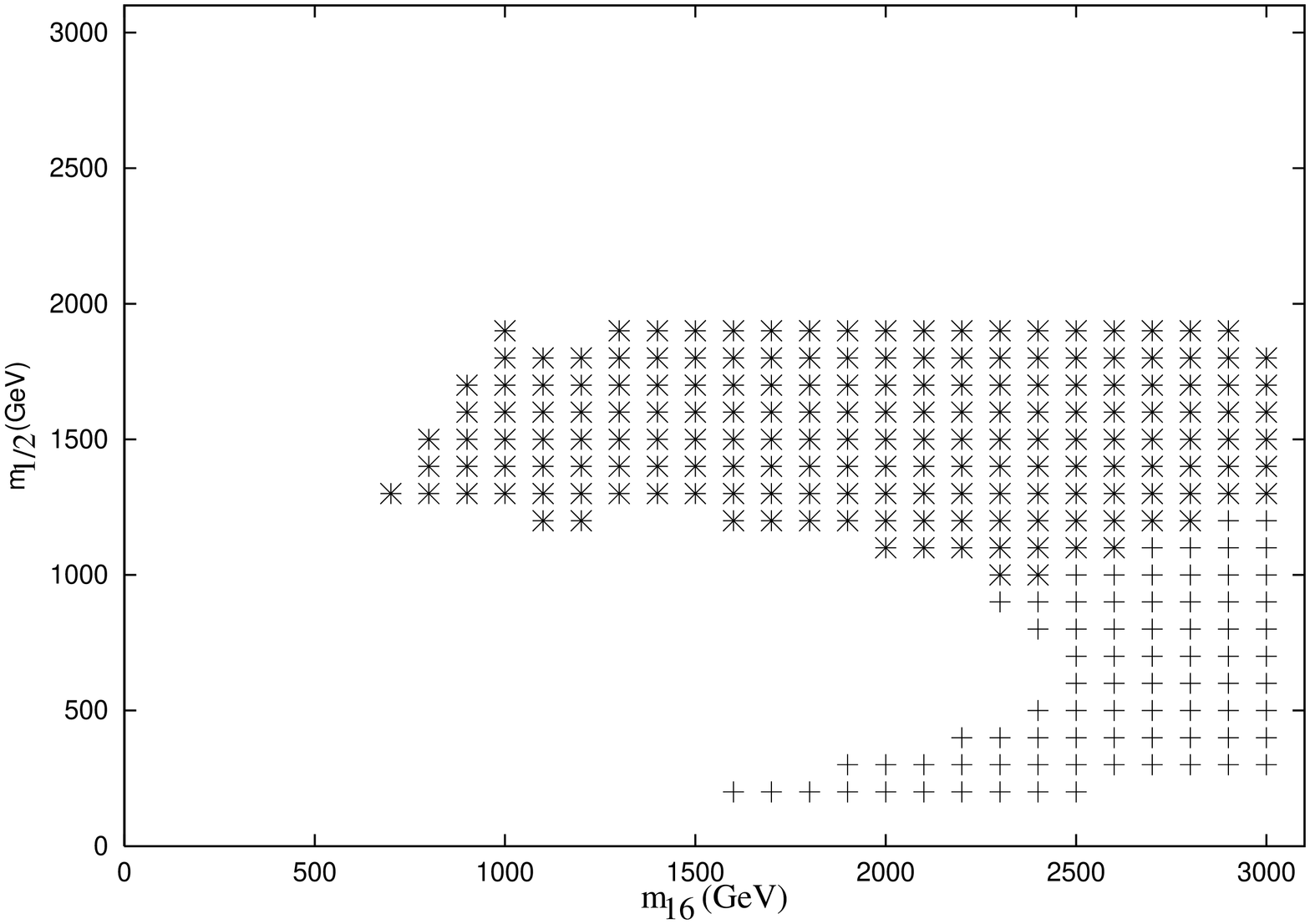,width=12cm,angle=270}
}
\caption{\sl{The same as fig. \ref{tbtaum0p1md5a0_1}, with $\MTN = .8\MSX$.
           }}
     \label{tbtaum0p.8md5a0_1}
\end{figure}

In fig. \ref{tbtaum0p1md5a0_1} we present the $\MSX - \MHF$
plane for $A_0 = -\MSX$ in the universal model. A large APS is obtained
by the unification criterion alone. 
For each $\MSX$ there are lower and upper bounds on $\MHF$. 
For $\MSX < 1200$GeV relatively low values of $\MHF$ are excluded by 
REWSB while very high values are excluded by the requirement that the 
neutralino be the LSP.
The value of $\MSX$ can be as low as 700 GeV, which corresponds 
$\MHF \geq$ 1100 GeV, yielding $\MGL \geq$ 2422 GeV, 
$\MSQ \approx$ 2200 GeV, $\MSL \approx$ 829 GeV.
For $\MSX \gapp 1200$GeV, 
low values of $\MHF$ are quite common.
Scanning over the APS we find that the lowest allowed gluino mass is
just above the experimental lower bound. 
Corresponding to this gluino mass the minimum sfermion masses are 
$\MSQ \approx$ 1200 GeV,
$\MSL \approx$ 1200 GeV.

As $A_0$ is further increased algebraically from $-\MSX$, the APS
slightly decreases due to  unification and REWSB constraints. 
For $A_0 = 0$, we obtain
an upper limit  $\MSX \leq$ 2400 GeV. However, the lower limits 
on $\MSX$ is relaxed by $\sim$ 200 GeV in comparison to the 
$A_0 = -\MSX$ case. 
As we further increase the value
of $A_0$ to $A_0 = \MSX$, the APS is almost the same as that for $A_0 = -\MSX$.
This trend is observed in all cases irrespective of
universality or non-universality of the scalar masses and even 
for $b-\tau$ unification. 
We find that as the absolute value of $A_0$ increases, Yukawa
unification is less restricted, while REWSB is somewhat disfavoured.
When both act in combination, we get a relatively large
APS for $\left |A_0\right | = \MSX$ and a somewhat smaller one
for $A_0 = 0$.

As the potential  constraints are switched on for $A_0 = -\MSX$, 
an interesting upper bound on $\MHF$ for each given $\MSX$ 
is imposed by the UFB1 constraint (fig. \ref{tbtaum0p1md5a0_1}). 
As a result practically   over the entire APS, 
the gauginos are required to be significantly lighter than the sfermions.
Moreover, the allowed gaugino masses are accessible to 
searches at the LHC. 

We next focus on the impact of a particular type of non-universality 
($\MTN <\MSX$) for the negative $A_0$ scenario. The shape of the APS 
is affected appreciably.
As $\MTN$ decreases, $Y_b$ gets larger SUSY threshold corrections
than $Y_{\tau}$ and $Y_t$; this disfavours  Yukawa unification.
On the other
hand $\MHU$ and $\MHD$ becomes more negative for even smaller values of
$\MSX$ and $\MHF$, which disfavors REWSB. The overall APS is somewhat smaller
compared to the  universal case, which is illustrated in fig. 
\ref{tbtaum0p.8md5a0_1} for $\MTN = .8\MSX$ (compare with fig. 
\ref{tbtaum0p1md5a0_1}).
The UFB1 constraint still imposes an upper bound on the gaugino mass
for a given $\MSX$ as in the universal case.
As a result the gauginos
are within the striking range of LHC practically over the entire APS.  
We also note from fig. 
\ref{tbtaum0p.8md5a0_1} that $\MSX \gapp 1600$ GeV over the entire APS. 

\begin{figure}[htb]
\centerline{
\psfig{file=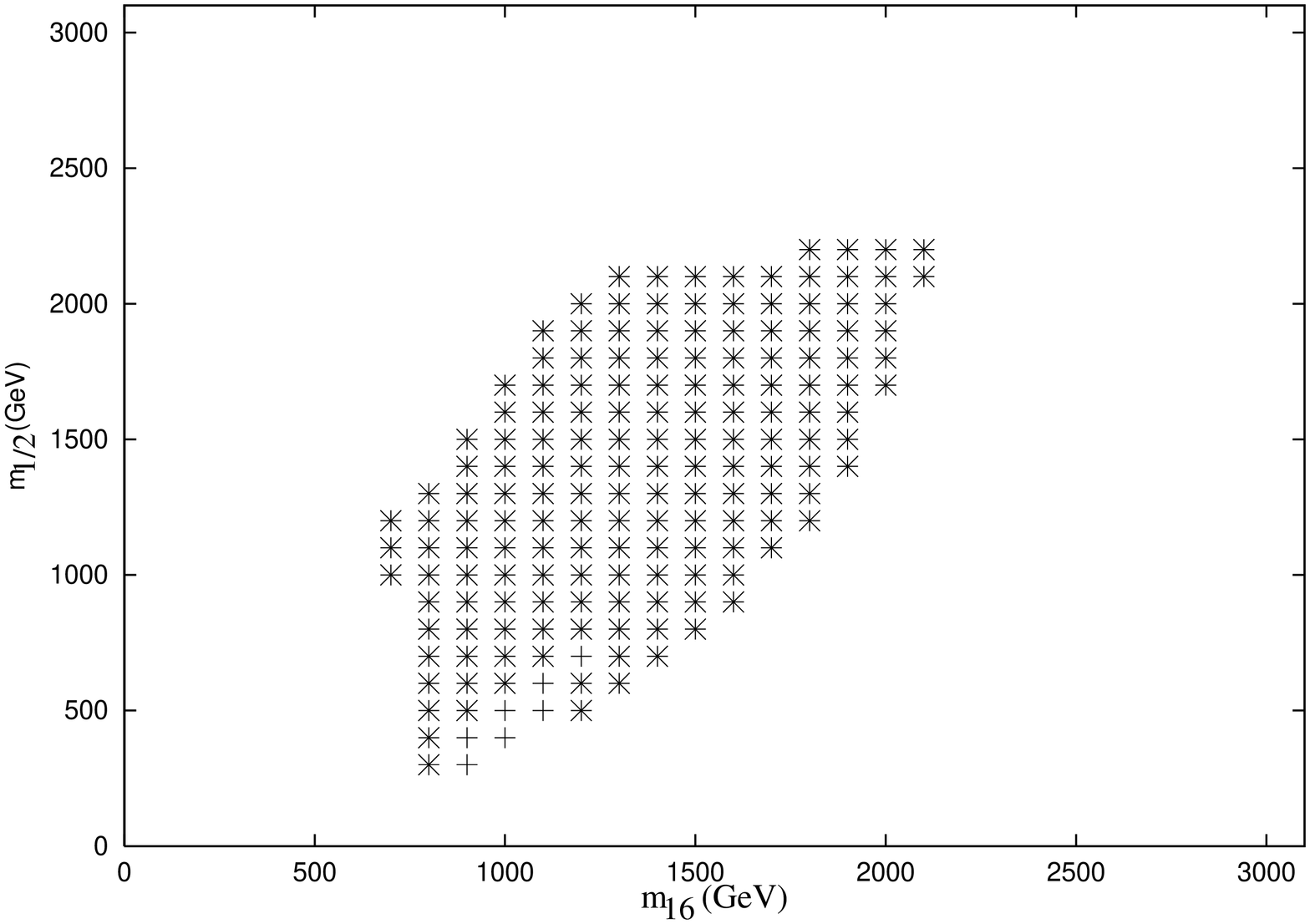,width=12cm,angle=270}
}
\caption{\sl{The same as Fig. \ref{tbtaum0p1md5a0_1}, with $\MTN = 1.2\MSX$.
           }}
     \label{tbtaum0p1.2md5a0_1}
\vspace*{5mm}
\centerline{
\psfig{file=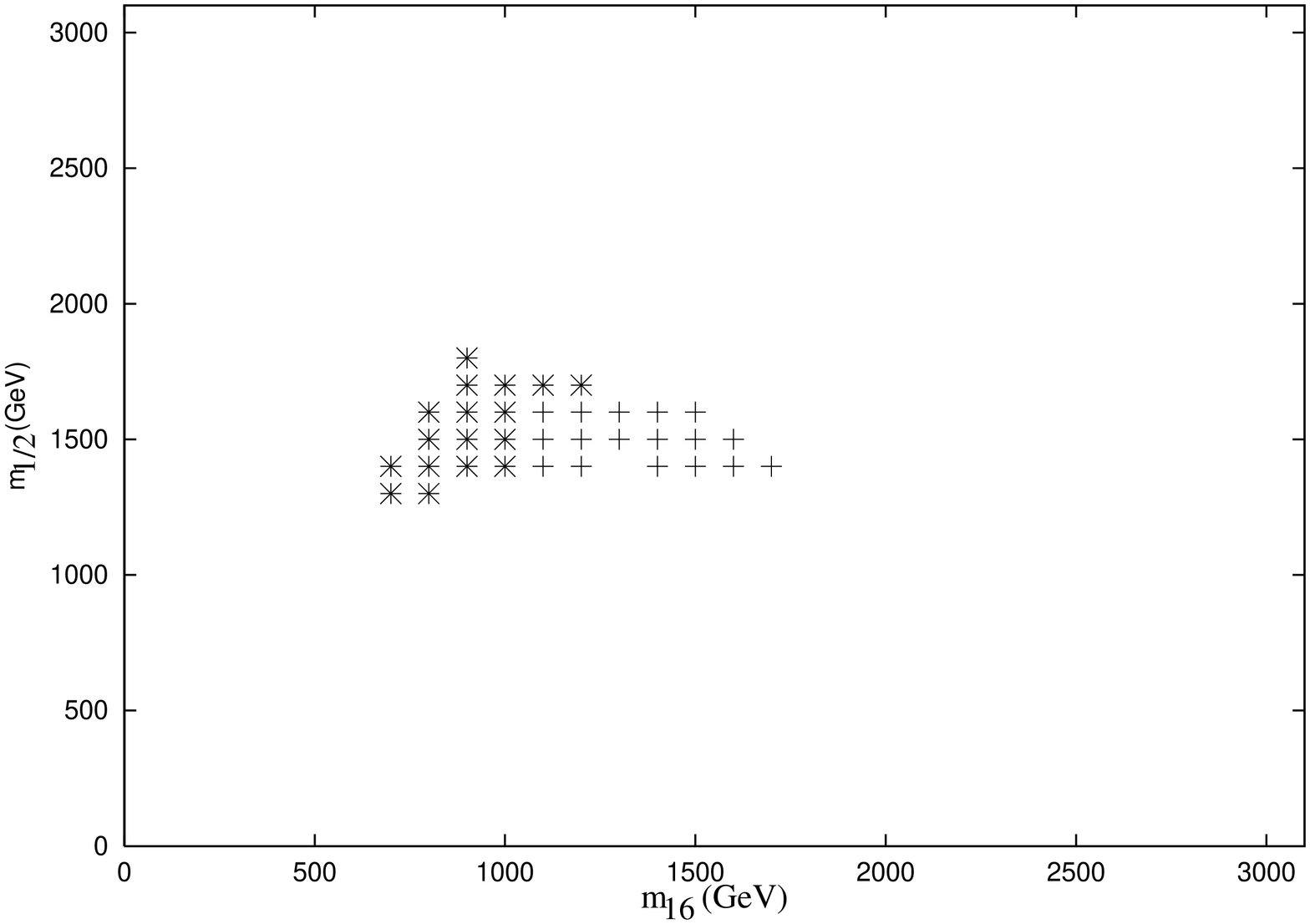,width=12cm,angle=270}
}
\caption{\sl{The same as fig. \ref{tbtaum0p1md5a0_1}, with $m_D = \MSX/3$.
           }}
     \label{tbtaum0p1md3a0_1}

\end{figure}

For a different pattern of non-universality ($\MTN > \MSX$), 
Yukawa unification alone  narrows down  the  APS considerably. However,
it is seen that regions with simultaneously low values of $\MSX$ and $\MHF$ 
are permitted. This happens in this specific nonuniversal scenario  only. 
On the otherhand, the parameter space with large $\MSX$ and small $\MHF$, 
preferred by the earlier scenarios, is disfavoured. 
With $\MTN = 1.2\MSX$ (fig. \ref{tbtaum0p1.2md5a0_1}), 
it is found that 
$\MHF\gapp$ 300 GeV. 
The unification allowed parameter space, however, is very sensitive to the
UFB conditions which practically rules out the entire APS for negative $A_0$.
No major change is noted in the APS for $A_0 = 0$ and $A_0 = \MSX$
apart from the fact that the UFB constraints get weaker.

We now discuss the impact of larger  D-terms on the parameter space.
For example, with  $m_D = \MSX/3$,  the APS reduces
drastically in the universal as well as 
non-universal scenario with $\MTN < \MSX$, irrespective of $A_0$. 
This is illustrated in fig. \ref{tbtaum0p1md3a0_1}. and  is in 
complete agreement with our qualitative discussion in the earlier
section.

Only in the specific nonuniversal scenario with $\MTN > \MSX$, 
slightly larger $m_D$ is preferred.
However, $m_D$ cannot be increased arbitrarily.
For  $\MTN=1.2\MSX$, the APS  begins to shrink again 
for  $m_D \gapp \MSX/3$ and  we find no allowed point for $m_D = \MSX/2$.
 
As $\MTN$ is increased  further, Yukawa unification  occurs in a 
narrower APS. This, nevertheless,  is a phenomenologically interesting 
region where  lower  $\MSX-\MHF$ values can be accommodated.
For example, $\MSX(\MHF) =$400(300)GeV is allowed with
 $\MTN=1.5\MSX, m_D=\MSX/3$, $A_0$=0 and tan$\beta \sim$ 51, 
leading to $\MGL$ = 742GeV, $\MSQ \approx$ 700 GeV, $\MSL \approx$ 400GeV 
and $m_{\tilde \tau_1}$ = 274GeV. 
However, we cannot increase $\MTN$ arbitrarily either, the APS reduces
drastically for $\MTN \gapp 1.5 \MSX$ irrespective of the
value of $m_D$. This trend qualitatively remains 
the same even if $A_0$ is changed. This effect can be seen in 
$b-\tau$ unification as well.                        

\subsection{$b$-$\tau$ Yukawa unification}
In our earlier work\cite{paper1} without D-terms, we had shown
that the APS is strongly restricted due to Yukawa unification
and UFB constraints. The minimum value of tan$\beta$ required for
unification is $\approx$ 30.
If D-terms are included, 
Yukawa unification and REWSB occur over a larger region of the
parameter space.
This is primarily due to two reasons: i) Yukawa unification can 
now be accommodated for lower values of tan$\beta$ ($\sim 20$) and ii) 
REWSB is allowed at somewhat higher values of $\tan\beta$ than 
the values permitted in $m_D = 0$ case.
This reduces the conflict between Yukawa unification and REWSB.
As a result $\MHF$ almost as low as that allowed by the LEP bound on the 
chargino mass is permitted over a wide range of $\MSX$. In some cases 
the upper bound on $\MSX$ for a given $\MHF$ is also relaxed.
Similarly  for a fixed $\MSX$, 
the upper bound on $\MHF$ is sometimes relaxed by few hundred GeVs.
Through out this work we require this  partial unification to an 
accuracy of $<$ 5\%.

\begin{figure}[htb]
\centerline{
\psfig{file=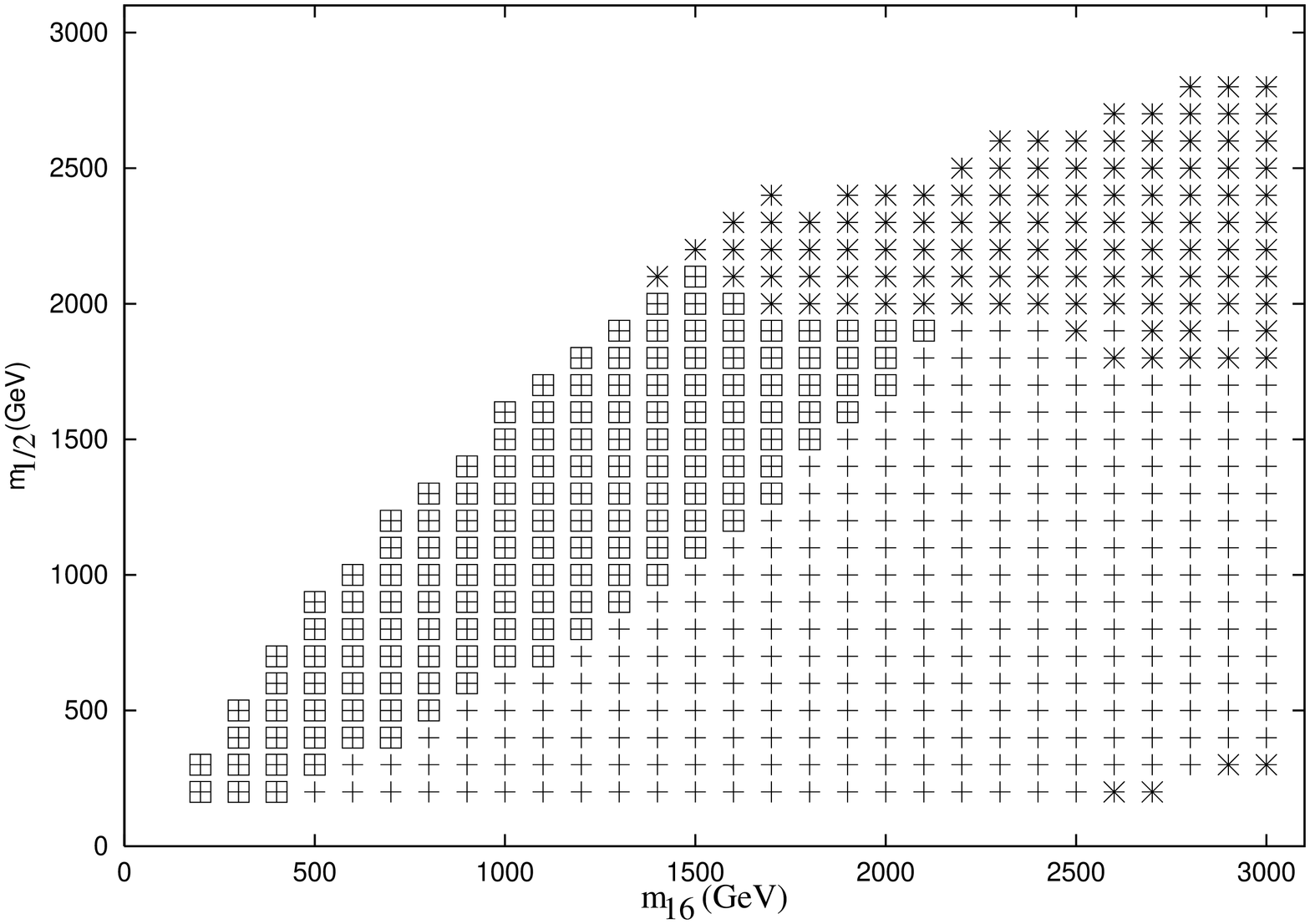,width=12cm,angle=270}
}
\caption{\sl{The allowed parameter space in the universal scenario with
             $b-\tau$ unification $\le$ 5\%. All the points are allowed 
             by the Yukawa unification criterion; the asterisks are 
             ruled out by UFB1 and the boxes are ruled out by UFB3.
             We set $m_D = \MSX/5$ and $A_0=-2m_{16}$.
            }}
    \label{btaum0p1md5a0_2}
\vspace*{5mm}
\centerline{
\psfig{file=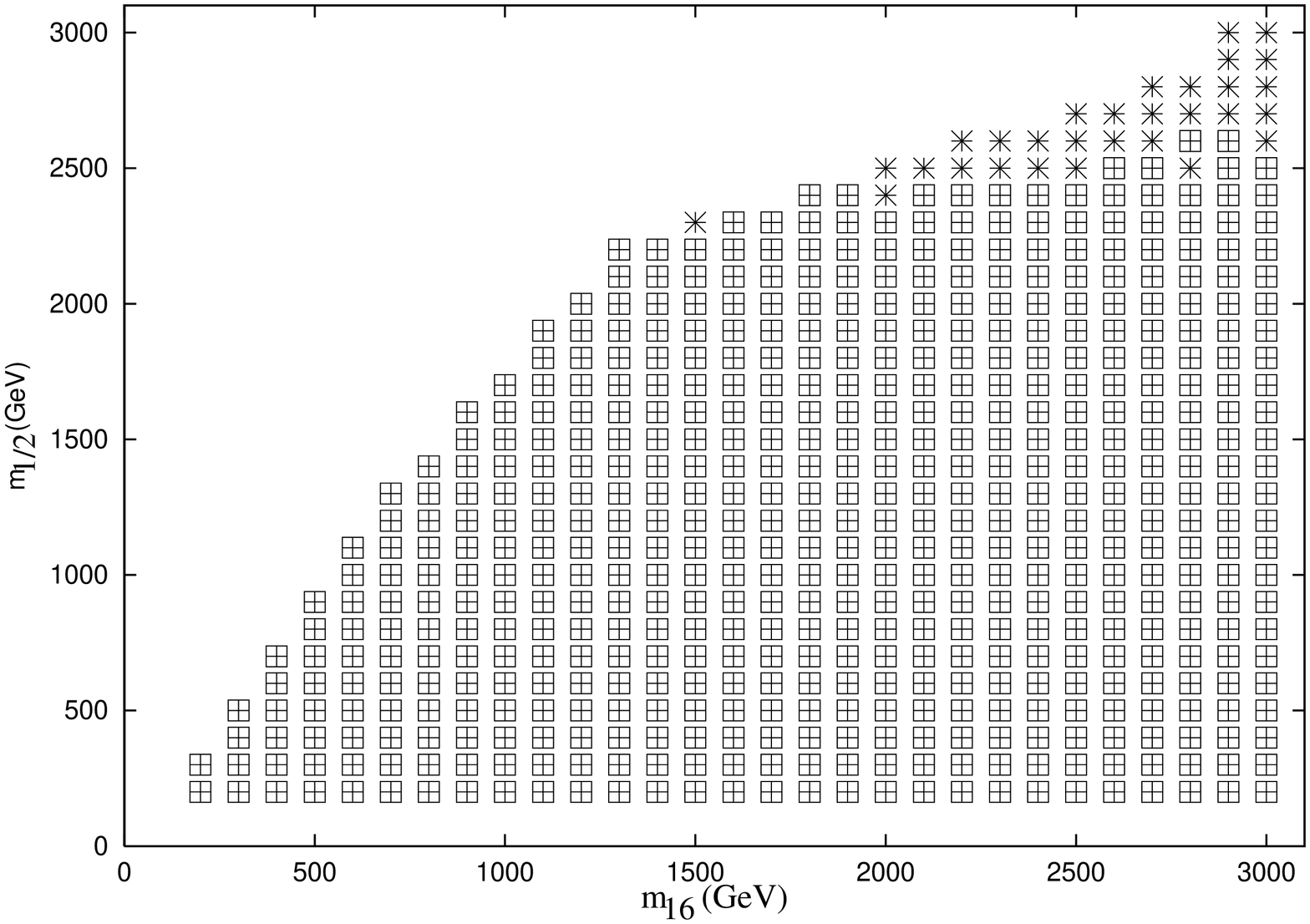,width=12cm,angle=270}
}
\caption{\sl{The same as Fig. \ref{btaum0p1md5a0_2}, with $\MTN = .6\MSX$.
           }}
     \label{btaum0p.6md5a0_2}
\end{figure}

Now we will focus our attention on large negative values  of 
$A_0$ ($A_0 = -2\MSX$) with $m_D = \MSX/5$ 
in the universal scenario. The unification allowed
APS, as shown in fig. \ref{btaum0p1md5a0_2}, expands 
compared to the $m_D$ = 0 scenario (compare with  fig. 6 of \cite{paper1}). 
Moreover, the  phenomenologically interesting scenario with light 
gauginos but very heavy sleptons and squarks beyond the reach of LHC,
which was rather disfavoured without the D-terms (see \cite{paper1}),  
is now viable.
Without the D-term, the APS was  severely restricted by the UFB conditions
for  large negative 
values of $A_0$. 
As discussed earlier,  inclusion  of the D-term increases the value 
of $\mu$. As a result  UFB1 looses its constraining power; lower values of
$\MHF$ are allowed for large $\MSX$ by UFB1. On the other hand 
as the value of D-term increases, UFB3 becomes more powerful and the
upper bounds on $\MHF$ for relatively low values of $\MSX$ get stronger  
(\eg for $\MSX$ = 600(1000)GeV, $\MHF <$ 300(600)GeV).
For $A_0 > 0$, the APS again expands. However, the UFB 
constraints are  found to be progressively weaker as $A_0$ is
increased from $A_0 = -2\MSX$. 

We next consider the non-unversal scenerio $\MTN \ne \MSX$. 
If we take $\MTN < \MSX$ (say, $\MTN=.6\MSX$) and  $m_D = \MSX/5 $, 
the unification allowed parameter space 
for $A_0 = -2\MSX$, as shown in fig. \ref{btaum0p.6md5a0_2}, 
is more or less the same as in the universal scenario.
The entire APS is, however, ruled out due to a very powerful 
constraint  obtained from the UFB3 condition. This conclusion obviously
holds for larger values of $m_D$.

\begin{figure}[htb]
\centerline{
\psfig{file=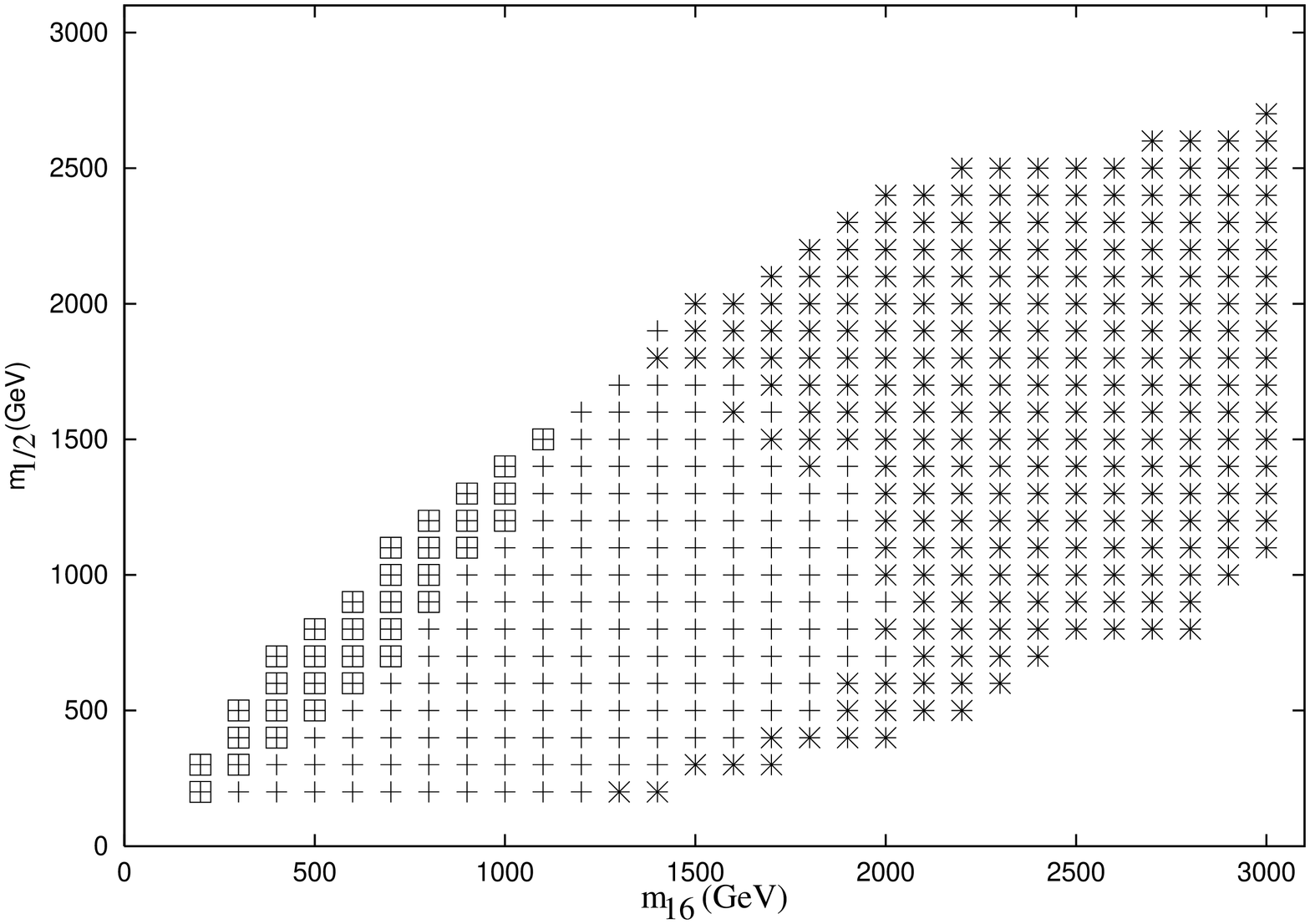,width=12cm,angle=270}
}
\caption{\sl{The same as Fig. \ref{btaum0p1md5a0_2}, with $\MTN = 1.2\MSX$.
           }}
     \label{btaum0p1.2md5a0_2}
\end{figure}

For $\MTN > \MSX$ and large negative $A_0$ ($A_0 = - 2\MSX$),
the unification allowed APS (fig. \ref{btaum0p1.2md5a0_2}) 
is smaller compared to that in the universal case 
(fig. \ref{btaum0p1md5a0_2}). The same trend was also observed 
with $m_D=0$\cite{paper1}. The APS, however, is significantly larger than 
that for $m_D = 0$. For a given $\MHF$ ($\MSX$) the upper-bound on $\MSX$
($\MHF$) gets weaker for non-zero D-terms. Relatively light 
gluinos consistent with current
bounds are allowed over a larger region of the parameter space.
The UFB constraints restrict the APS further and put rather strong bounds
on $\MHF$ and $\MSX$. A large fraction of this restricted APS
is accessible to tests at LHC energies. 
The usual reduction of the APS due to unification constraints as 
$A_0$ is increased from $A_0 = - 2\MSX$ also holds in this nonuniversal 
scenario.

If we increase $m_D$ further, the APS due to Yukawa 
unification reduces for reasons already discussed. 
The UFB1 constraint also gets weaker. On the other hand 
the UFB3 constraints  become rather potent.
For example, i) with $A_0 = - 2\MSX$ and $m_D \gapp \MSX/3$ the entire APS
for $\MTN = \MSX$ or $\MTN < \MSX$ is ruled out. ii) $A_0 = -\MSX$ 
and $m_D \gapp \MSX/2$ the entire APS corresponding to $\MTN = \MSX$ 
or $\MTN < \MSX$ is ruled out. On the other hand, for $\MTN > \MSX$ 
the APS further reduces as $m_D$ is increased.

\section{Conclusion}
For moderate values of the D-terms $(m_D \approx \MSX/5)$,
the APS expands in general compared to the $m_D=0$ case for both 
$b-\tau$ and $t-b-\tau$ Yukawa unification. 
A large fraction of the enlarged APS is, however, reduced by the 
requirement of vacuum stability and the predictive power is not lost
altogether. D-terms with much larger magnitudes, however, are not 
favourable for unification. 
In the $t-b-\tau$ Yukawa unified model (accuracy $\leq 5\%$), a band of
very low gaugino mass close to the current experimental lower limit 
is a common feature in the presence of D-terms. 
For a given $\MSX$ there is an upper bound on $\MHF$ from unification 
and stability of the potential constraints.
This happens for 
$-\MSX\lapp A_0\lapp\MSX$. Outside this range of
$A_0$, the APS is very small with  sparticle masses of the first 
two generations well above 1 TeV. In $b-\tau$ unification, 
UFB3 strongly
restricts the APS while UFB1 becomes less potent in the presence of 
D-terms.  

{\em \bf Acknowledgements}:
The work of AD was supported by DST, India (Project No.\ SP/S2/k01/97)
and BRNS, India (Project No.\ 37/4/97 - R \& D II/474).
AS acknowledges CSIR, India, for his research fellowship.

\vspace*{5mm}


\end{document}